\documentclass[12pt,psfig,epsfig]{article}
\usepackage{graphicx}
\usepackage{dcolumn}
\usepackage{bm}
\usepackage{epsfig}
\newcommand{\beq}{\begin{equation}}
\newcommand{\eeq}{\end{equation}}
\newcommand{\bea}{\begin{eqnarray}}
\newcommand{\eea}{\end{eqnarray}}
\begin{document}

\begin{center}{PeV Gamma Rays from Interactions of Ultra High Energy Cosmic Rays
 in the Milky Way}\\ 
\medskip
{Nayantara Gupta\footnote{Email: nayan@rri.res.in}}\\

{Department of Physics, 
Indian Institute of Space Science and Technology, Trivandrum, INDIA\\
Astronomy \& Astrophysics Group, Raman Research Institute, Bangalore, INDIA}
\end{center}
\begin{abstract}
The PeV gamma ray background produced in the interactions of ultra high energy cosmic rays with the ambient matter and radiations during their propagation in the Milky Way has been calculated in this paper. If the primary ultra high energy cosmic rays are produced from Galactic point sources then those point sources are also emitting PeV gamma rays. We discuss that the detection of galactocentric PeV gamma rays in the future would be a signature of the presence of EeV cosmic accelerators in the Milky Way.
 
\end{abstract}

PACS numbers: 95.85.Pw, 98.35.Jk, 98.35.Gi

\date{\today}

\section{Introduction}
The observational data from extensive air shower experiments like Hires \cite{hires}, AGASA \cite{agasa}, Yakutsk \cite{yak} and Auger \cite{auger} have been
 widely used to undestand the origin of ultra high energy cosmic rays (UHECRs) \cite{nagano,bhat,olinto}. The maximum energy of cosmic rays $E_{max}$ from a cosmic accelerator is determined by the size (R), magnetic field (B) of the acceleration region, shock velocity ($\beta$) and charge of the cosmic ray particle (Z), $E_{max}=Ze\beta BR$ \cite{hillas}.  
It is not yet known what could be the maximum energy of the cosmic rays originated from sources inside Milky Way. Most of the cosmic ray sources in the Milky Way are located near its center, as a result a galactocentric anisotropy is expected in the UHECR spectrum. The observed UHECR spectrum does not show such anisotropy upto 1EeV energy. The properties of cosmic rays in the energy range of $0.1$PeV and 1EeV have been studied by KASCADE-Grande experiment \cite{kas1}.
It has been shown earlier that the knee like structure near 4 PeV in the all particle spectrum is due to the decrease in the flux of light nuclei \cite{antoni,apel1}. The all particle energy spectrum of cosmic rays reconstructed from the KASCADE-Grande data using three different techniques and the hadronic interaction models QGSJET II, FLUKA shows concavity and a small discontinuity near 0.01 EeV, 0.1 EeV respectively \cite{kas2}. The energy dependence of composition of UHECRs can not be determined directly as their flux decreases with increasing energy. It is determined indirectly by observing extensive air showers in the atmosphere. The average depth of the shower maximum depends on the energy and mass of the primary particle. The current observational results tell us above 1EeV there is a  gradual increase in average mass of cosmic rays upto energy 59EeV \cite{abraham}.  Recently it has been discussed \cite{calvez} that cosmic ray nuclei of energy 1EeV to 10EeV can be originated from old gamma ray bursts or supernova explosions in our Galaxy and their directions may become isotropic as they spend long time in the micro-Gauss magnetic field of Galaxy. The inhomogeneous magnetic field in our Galaxy is responsible for trapping the charged particles for long time. The field strength has been estimated to be between $60 \mu G$ and $400 \mu G$ in the inner region of the Galaxy \cite{crock}. 
Inside Milky Way there are supernova remnants (SNRs), pulsar wind nebulae (PWN), microquasars and many unidentified sources of $\gamma$ rays. The TeV gamma ray spectrum from some of the SNRs in our Galaxy may be explained with hadronic interaction models and inverse Compton emission by electrons 
\cite{tycho,sn1006}. SNRs have long been identified as attractive sites for cosmic ray accelerations. In particular, near the Galactic center (GC) region, there is a large population of $\gamma$ ray sources. The high energy photon fluxes emitted by them have been observed by HESS \cite{hess}, FERMI \cite{fermi}, AGILE \cite{agile}, MILAGRO \cite{milagro} and other gamma ray detectors. The photon flux from GC region as observed by FERMI-LAT has been fitted with conventional theoretical model of cosmic ray interactions with ambient gas medium, inverse Compton scattering of electrons and positrons by interstellar radiations\cite{strong1}. If one considers only the protons, then $pp$ and $p\gamma$ interactions 
are the possible mechanisms of $\gamma$ ray production. The cross-section of $pp$ interactions is much higher that $p\gamma$ interactions \cite{pdr}. In the GC region the radiation and interstellar gas densities are high compared to the outer regions of Galaxy. The Galactic interstellar radiation field has been calculated by Moskalenko et al. \cite{mos1}. With the average number density of hydrogen gas molecules $n=10 cm^{-3}$ and $120 cm^{-3}$ \cite{crock} and photon density of Galactic radiation field \cite{mos1}, one finds $pp$ process is much more important than $p\gamma$ process in the central region of Milky Way. In these interactions nearly $20\%$ of the original cosmic ray proton's energy goes to the neutral pion and it decays to two high energy gamma rays of equal energy. Hence, we expect 100 PeV gamma rays from 1EeV cosmic ray protons.
1 EeV proton of extra-galactic origin may appear like UHECR of Galactic origin in the high magnetic field of GC region. If the UHECRs are protons and they are of Galactic origin then a galactocentric flux of secondary PeV gamma rays may be expected in interactions of these protons with ambient matter.    
The UHECR heavy nuclei would be deflected more in Galactic magnetic fields compared to the protons. They would interact with ambient hydrogen molecules producing pions, subsequently decaying to very high energy photons and neutrinos. Also there may be photo-disintegration of heavy nuclei and photo-pion production by heavy nuclei. In this paper we calculate the secondary PeV-EeV gamma ray flux produced by the interactions of diffuse UHECRs with ambient matter and radiations in the central region of Milky Way. 
 The secondary PeV gamma rays/neutrinos may be anisotropic because of the strong magnetic field in the GC region. Moreover, if the UHECRs are produced by point sources in the central region of Milky Way then we expect to detect a flux of galactocentric PeV gamma rays from those point sources, above the PeV gamma ray background calculated in this work. Thus the detection of PeV gamma rays would be useful to know the maximum energy of the cosmic rays produced by sources in the Milky Way.      

\section{Pure Hadron Interactions}
The cosmic ray flux emitted from the sources can be assumed to be a mixture of light and heavy nuclei. They interact with the matter molecules in the Milky Way during propagation producing secondary gamma rays and neutrinos. The cosmic ray protons' interactions with ambient matter protons can be written as, $pp\rightarrow n_{\pi}\pi^{\pm,0}+X$, where $n_{\pi}$ is the pion multiplicity. The charged pions decay to high energy leptons and the neutral pions to high energy gamma rays. 
The secondary particles production in $pp$ interactions has been studied earlier in \cite{vb1,vb2,mfat,rm1,grasso1}. 

The interaction of cosmic ray nuclei with cold ambient matter leads to the production of neutral pions, this pion flux has been calculated in \cite{steck1}. The neutral pions subsequently decay to energetic gamma rays. The emissivity of gamma rays has been calculated in \cite{anch1}.
The UHE photon flux produced in $pp$ interactions and subsequent $\pi^0$ decay is
\beq
F_{\gamma,pp}(E_{\gamma})= \frac{t_{esc,p}}{t_{pp}}G_{pp}(E_{\gamma})Y_{\alpha},
\label {gamma_p}
\eeq
where $E_{\gamma}=0.1 E_p$. $t_{esc,p}$ and $t_{pp}\simeq 10^{15}/n_H sec$ are the escape and $pp$ interaction time scales for $\pi^0$ production of cosmic ray protons respectively, $n_H$ is number density of ambient hydrogen molecules in the Milky Way. The cross section for $pp$ interactions is included in $t_{pp}$. The proton flux is contained in this term
\beq
G_{pp}(E_{\gamma})=2\int_{E_{\pi^0,min}}^{E_{\pi^0,max}}\frac{dN_p(E_{\pi^0})}{dE_{\pi^0}}\frac{dE_{\pi^0}}{(E_{\pi^0}^2-m_{\pi^0}^2)^{1/2}},
\label{g_pp}
\eeq
where $\frac{dN_p(E_{\pi^0})}{dE_{\pi^0}}=A_p E_{\pi^0}^{-\alpha}$, $A_p$ is normalisation constant and $\alpha$ is spectral index of the proton spectrum, $E_{\pi^0,min}=E_{\gamma}+m_{\pi^0}^2/(4E_{\gamma})$. The maximum energy of pions is the maximum energy of the cosmic ray proton/nucleon $E_{\pi^0,max}=E_n^{max}$.
  In $pp$ interactions $\pi^0$ carries on the average $20\%$ of the cosmic ray proton's energy. The spectrum-weighted moments $Y_{\alpha}$ has been calculated from \cite{anch1}
\beq
Y_{\alpha}=\int_0^1 x^{\alpha-2}f_{\pi^0}(x) dx.
\label{y_l}
\eeq
The function $f_{\pi^0}(x)\simeq8.18x^{1/2}\Big(\frac{1-x^{1/2}}{1+1.33x^{1/2}(1-x^{1/2})})\Big)^4\Big(\frac{1}{1-x^{1/2}}+\frac{1.33(1-2x^{1/2})}{1+1.33x^{1/2}(1-x^{1/2})}\Big)$ with $x=E_{\pi^0}/E_p$.  
For $\alpha=2.2,3$ we find $Y_{\alpha}$ is 0.079 and 0.02 respectively.
The gamma ray flux $F_{\gamma,Ap}(E_{\gamma})$ produced in pure hadronic interactions of cosmic ray nuclei has been calculated with this expression. 
\beq
F_{\gamma,Ap}(E_{\gamma})=\frac{t_{esc,A}}{t_{hadr}}G_{Ap}(E_{\gamma})Y_{\alpha},
\label{gamma_fe}
\eeq
where $t_{esc,A}$ is the escape time scale of cosmic ray nuclei of mass number $A$. The flux of iron nuclei can be expressed as a power law in energy similar to the proton flux $\frac{dN_{FE}(E_{FE})}{E_{FE}}=A_{FE}E_{FE}^{-\alpha}$.
  The hadronic interaction time scale of cosmic ray nuclei can be estimated with hadronic interaction cross section for $\pi^0$ production $\sigma_{hadr}^A=34.6A^{3/4}mb$ \cite{leb}, where $A$ is mass number of nuclei. This time scale is $t_{hadr}\simeq A^{-3/4}/n_H\times10^{15}sec$.
The expression for $G_{Ap}$ for iron nuclei can be calculated similar to $G_{pp}$, except the proton flux is to be replaced by flux of iron nuclei per unit nucleon energy. If the proton and iron nuclei flux normalisation constants are equal ($A_p=A_{FE}$) then using $E_{FE}=56 E_p$ one gets
\beq
G_{Ap}(E_{\gamma})=56^{-\alpha+1}G_{pp}(E_{\gamma}).
\label{g_ap}
\eeq
The escape time scales of UHE cosmic ray protons and nuclei are not known. In principle one may vary the escape time scale and source spectrum of cosmic rays to fit the observed gamma ray flux. 
The ratio of the fluxes of gamma rays and cosmic ray nuclei of mass number $A$ at the same energy can be expressed as
\beq
\frac{F_{\gamma,Ap}}{F_{CR,A}}=\frac{t_{esc,A}}{t_{hadr}}\frac{2 Y_{\alpha}}{\alpha} A^{-\alpha+1}
\label{r_fluxes}
\eeq
 
The secondary nucleons produced in pure hadron processes may also interact with ambient matter and radiations to produce high energy gamma rays and neutrinos.
\section{Photo-Hadron Interactions}
The cosmic ray nuclei are assumed to be propagating through an isotropic photon background with energy $\epsilon$ and photon density per unit energy $n(\epsilon)$.
The energy of a cosmic ray nucleus $E_{tot,n}=\gamma_n A m_n$ where $\gamma_n$ is the Lorentz factor of the cosmic ray nucleons, $A$ is mass number and $m_n$ is the mass of each nucleon in the nucleus.
The time scale of photo-disintegration of cosmic ray nuclei in the radiation field of Milky Way is $t_{phot-d}=\frac{\lambda_A}{c}$ where $\lambda_A$ is the mean free path of photo-disintegration of cosmic ray nucleus \cite{anch1}
\beq
\frac{1}{t_{phot-d}}=\frac{c}{\lambda_A}=\frac{c}{2}\int_0^{\infty}\frac{n(\epsilon)}{\gamma_n^2 \epsilon^2} d\epsilon \int_0^{2\gamma_n \epsilon} \epsilon^{\prime} \sigma_A(\epsilon^{\prime})d\epsilon^{\prime}.
\label{r_phot-d}
\eeq
The cross section for photo-disintegration of a nucleus of mass number $A$ by a photon of energy $\epsilon^{\prime}$ in the rest frame of the nucleus is $\sigma_A(\epsilon^{\prime})$. Detailed discussions on cross sections for all different nuclear species can be found in \cite{steck2}. 
Eqn.(\ref{r_phot-d}) can be approximated as
\beq
\frac{1}{t_{phot-d}}=\frac{\pi \sigma_0\epsilon'_0 \Gamma}{4 \gamma_n^2}\int_{\epsilon'_0/2\gamma_n}^{\infty} \frac{d\epsilon}{\epsilon^2}n(\epsilon).
\label{fig_phot-d}
\eeq 
$\Gamma=8 MeV$ is the width, $\epsilon'_0=42.65 A^{-0.21} MeV$, for $A>4$ is the central value of the GDR energy band and $\sigma_0=1.45\times10^{-27}A$ $cm^2$. 
The infrared radiation field in our Galaxy at different distances from the GC region has been calculated by Moskalenko et al. \cite{mos1}.
 We have approximated the infrared photon number density per unit energy as $n(\epsilon)=\frac{1}{\epsilon^2}$ $eV^{-1} cm^{-3}$ to calculate $t_{phot-d}$. The photo-disintegration time scale and the hadronic interaction time scale of iron nuclei are shown in Fig.1 with solid and dashed lines respectively. Below 100 EeV hadronic interaction is the dominant process of energy loss by iron nuclei. Moreover, the gamma rays produced in photo-disintegration of nuclei have much lower energy than the parent nuclei, hence it does not contribute significantly to the PeV gamma ray flux.
There can be photo-pion production by nuclei of mass number $A$ in the radiation field of Milky Way $A\gamma\rightarrow n_{\pi}\pi^{+,0}+X$. 
The iron nuclei of energy in the range of PeV-EeV satisfy the threshold energy condition for $A\gamma$ interactions with the x-ray background \cite{milk_x}. We find that the time scale of this interaction is much larger compared to the time scales of hadronic and photo-disintegration processes.
We have calculated the time scale of photo-pion production in $p\gamma$ interactions in the radiation field of GC region with photon number density $n(\epsilon)=\frac{1}{\epsilon^2}$$ eV^{-1} cm^{-3}$ and found it is much larger compared to the $pp$ interaction time scale with ambient hydrogen molecule density $n_H=10$$ cm^{-3}$. 
  
\begin{figure}
\centerline{\includegraphics{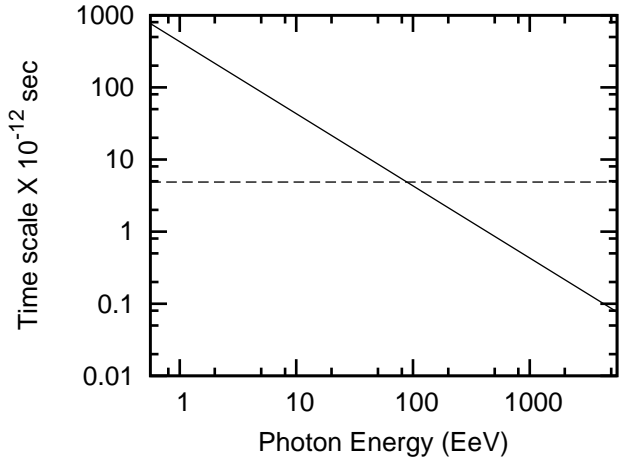}}
\caption{The time scales of photo-disintegration and hadronic interactions of iron nuclei with ambient hydrogen molecule density $n_H=10cm^{-3}$ are shown by solid and dashed lines respectively.}
\label{time}
\end{figure}
\section{Cosmic Microwave, Galactic and Extragalactic Radio Background}
 The secondary photons of energy 1 PeV satisfy the threshold energy condition
for production of $e^{\pm}$ pairs with cosmic microwave background (CMB) photons. Higher energy photons may interact with both cosmic microwave and radio background photons. The interaction of UHE photons ($E>1PeV$) with CMB is suppressed due to Klein Nishina suppression of interaction cross section. The Galactic and extra-galactic radio backgrounds have been measured by ARCADE 2 (Absolute Radiometer for Cosmology, Astrophysics, and Diffuse Emission) instrument \cite{arc}. This instrument measures the absolute temperature of the sky in search of distortions from a blackbody spectrum. It operates at centimeter wavelengths between fullsky surveys at radio frequencies below 3 GHz and the Far Infrared Absolute Spectrophotometer (FIRAS) survey at frequencies above 60 GHz. The sky temperatures measured by ARCADE 2 includes contributions from Galactic, extra-galactic and cosmic sources. The 2.7 K cosmic microwave background dominates the measured temperatures. Galactic emission is dominated by synchrotron emission with a smaller contribution from free-free sources. The integrated contribution of similar synchrotron and free-free emission in external galaxies constitutes an isotropic  extra-galactic radio background. Although, the CMB may be distinguished from Galactic or extra-galactic radio emission by the different frequency dependences, spectral fitting alone can not distinguish Galactic emission from an extra-galactic component of similar spectral behavior.
The method of determining the Galactic emission by using the north and south polar caps as convenient reference line of sights has been disussed in \cite{ar_g}. 
The measurement of radio emission from Galactic polar cap regions can be fitted with a single power law over the frequency range of 22 MHz to 10 GHz
\beq
T_{Gal}(\nu)=T_0(\nu/\nu_0)^{\beta},
\eeq
with spectral index $\beta=-2.55\pm 0.03$ and amplitude $T_0=0.498\pm 0.028$ K at reference frequency $\nu_0=1$ GHz.
The total Galactic emission can be estimated with a simple plane-parallel structure of Galaxy and a cosecant fit in Galactic latitude angle.   
The ARCADE 2 data show a rise in temperature of $50\pm 7$ mK at 3.3 GHz in addition to a CMB temperature of $2.730\pm 0.004$ K due to the extra-galactic background. The extr-agalctic radio background temperature has been measured in \cite{ar_e}. A power law fit to parametrise the extra-galactic background gives
\beq
T_{CRB}(\nu)=T_{CMB}+A(\nu/\nu_0)^{\beta},
\eeq  
where $T_{CMB}$ is the CMB baseline temperature $2.725\pm0.001$ K, $A=1.26\pm 0.09$ K is the power law amplitude at 1 GHz, $\beta=-2.6\pm 0.04$ is the power law index,in the frequency range of 22 MHz to 10 GHz.
The secondary UHE photons of energy 100 PeV produced in $\pi^0$ decay, satisfy the threshold condition for pair production with the radio photons of frequency 0.6 GHz. For the lower energy UHE photons energy loss due to Galactic radio background is not important. 
 Radio Air-Shower Test Array (RASTA) would be an extension of IceCube detector at the South Pole to detect air showers with surface array of radio antennas \cite{ice,bose}. UHE gamma rays may be detected in future with this detector. Photon induced showers will show a hundred times lower muon flux than hadron induced showers.  
Moreover, hadron induced air showers are more energetic and IceCube will have a nearly $100\%$ efficiency of detecing showers of energy more than 1 PeV if the shower axis intersects the geometric volume of the detector. This can be used to discriminate hadron induced showers from photon induced showers. 
 
PeV photons of extra-galactic origin are unlikely to reach us due to their cosmological distances and higher absorption by background radiations. 
Thus the observation of PeV gamma rays would provide an exclusive window to study the interactions of UHECRs in the Milky Way.
\section{Optical Depths in the Galaxy}
The optical depths for very high energy photons due to background radiations (radio to x-rays) were calculated by Gould $\&$ Schr\'eder \cite{gs}. 
The mean free path lenghts for interactions of UHE photons with extra-galctic radio background down to KHz frequency was calculated by Protheroe $\&$ Biermann \cite{pb}. We use the power law fits of sky temperatures due to the Galactic and extra-galctic radio backgrounds in the frequency range of 22 MHz to 10 GHz given in \cite{ar_g,ar_e} to calculate the mean free path lengths for pair($e^{\pm}$) production by very high energy gamma rays. The energy density spectrum of Galactic and extra-galactic radio backgrounds is
\beq
\frac{4\pi}{c}\nu[B_{CRB}(\nu)+B_{Gal}(\nu)]=\frac{8\pi k \nu^3}{c^{3}}(T_{CRB}(\nu)+T_{Gal}(\nu)).
\eeq
The number density of radio photons per unit energy is
\beq
\frac{dn({\epsilon})}{d\epsilon}=\frac{4\pi}{hc}\frac{B_{CRB}(\epsilon)+B_{Gal}(\epsilon)}{\epsilon}.
\eeq
The mean free path of pair production has been calculated following the formalism discussed in \cite{nayan1}. The threshold frequency of radio photons for pair production with 10 EeV gamma rays is nearly 6 MHz. We have assumed the power law fits of sky temperatures in frequency is valid upto this frequency.  
\begin{figure}
\centerline{\includegraphics{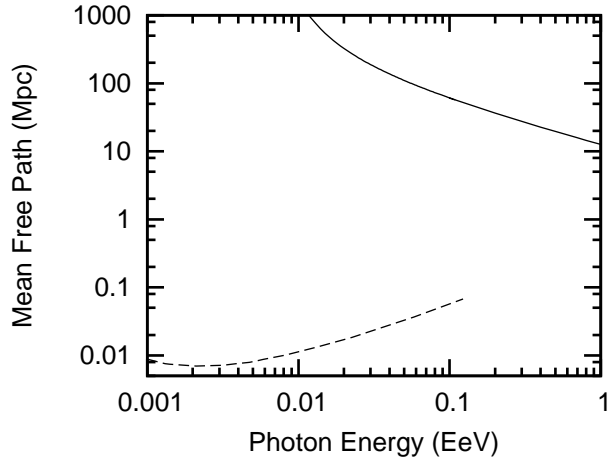}}
\caption{ Mean free path of pair production, dashed line: CMB, IR from \cite{mos1}; solid line: radio background (Galactic and extra-galactic) calculated using the radio background measured in \cite{ar_g,ar_e}}
\label{optd}
\end{figure}
Our result for the radio background is shown in Fig.\ref{optd}. It is comparable to the result given in \cite{pb}.
The mean free path length for pair production due to CMB is 10 Kpc for 1 PeV gamma rays \cite{pb}, which is nearly equal to the distance of GC from us $(\sim$ 8Kpc). For higher energy gamma rays it increases.
 
\section{PeV Gamma Ray Event Rates} 
\begin{figure}
\centerline{\includegraphics{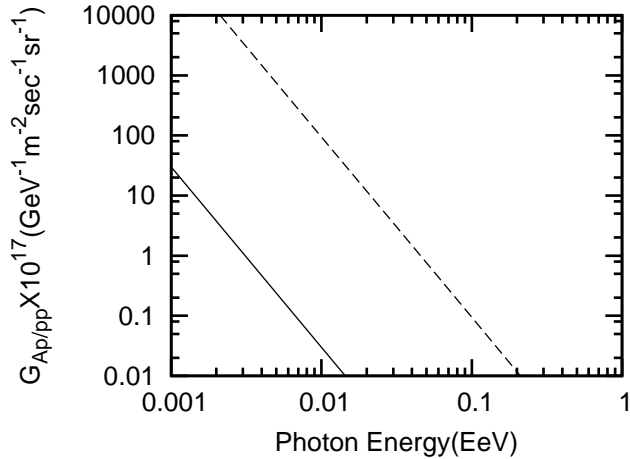}}
\caption{$G_{Ap}(E_{\gamma})$ and $G_{pp}(E_{\gamma})$ plotted with solid and dashed lines respectively for $A_{Fe}=A_p=1.45\times10^{6}GeV^{-1} m^{-2} sec^{-1} sr^{-1}$ and $\alpha=3$.}
\label{g}
\end{figure}

\begin{figure} 
\centerline{\includegraphics{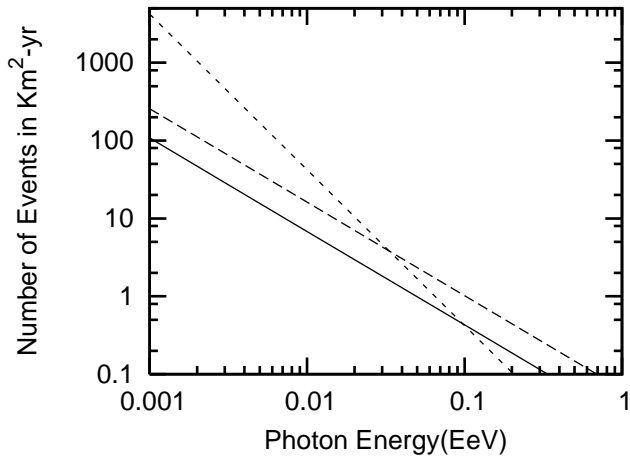}}
\caption{High energy gamma ray event rates in $Km^2-yr$, solid line: source spectrum only iron with $A_{Fe}=1 GeV^{-1} m^{-2} sec^{-1} sr^{-1}$, $\alpha=2.2$; dashed line: iron and proton $A_{Fe}=A_p=1 GeV^{-1} m^{-2} sec^{-1} sr^{-1}$, $\alpha=2.2$; dotted line: iron and proton $A_{Fe}=A_p=1.45\times 10^{6} GeV^{-1} m^{-2} sec^{-1} sr ^{-1}$, $\alpha=3.$}
\label{event}
\end{figure}

We find that pure hadronic interactions are more important compared to photohadronic processes in our study. In Fig.\ref{time} the hadronic interaction and photo-disintegration time scales are shown for iron nuclei with dashed and solid lines respectively.
The functions $G_{pp}(E_{\gamma})$, $G_{Ap}(E_{\gamma})$ in eqn.(\ref{g_pp}) and eqn.(\ref{g_ap}) are plotted in Fig.\ref{g}. 
The upper limit on the ratio of gamma ray to normal cosmic ray fluxes at the same energy was set at $2.4\times10^{-5}$ at 310 TeV with $90\%$ confidence limit by CASA-MIA experiment \cite{bor}. The spectrum weighted moment in eqn.(\ref{y_l})
$Y_{\alpha}=0.079, 0.02$ for spectral index $\alpha=2.2,3$ respectively.
If we apply the limit from CASA-MIA experiment in the case of PeV gamma rays then one has to assume that the ratio of escape and hadronic interaction time scales of iron nuclei is $\sim 0.03,5.6$ for $\alpha=2.2,3$ from eqn.(\ref{r_fluxes}) assuming all the gamma rays are produced by hadronic interaction of UHECR iron nuclei. If we assume all the gamma rays are produced by UHECR protons with spectral index $\alpha=2.2,3$ then the ratio of escape and $pp$ interaction time scales of protons is $0.33\times10^{-3},1.8\times10^{-3}$ respectively.
The high energy gamma ray event rates have been calculated assuming only iron nuclei at the source with source spectrum $\frac{dN_A(E_{Fe})}{dE_{Fe}}=A_{Fe} E_{Fe}^{-\alpha}$, $A_{Fe}=1 GeV^{-1} m^{-2} sec^{-1} sr^{-1}$ and spectral index $\alpha=2.2$. For $Km^2$ area detector the number of events expected in one year have been plotted against energy, shown with solid line in Fig.\ref{event}. We have not included the efficiency of the detector in this work. High energy photon absorption by CMB background at 2 PeV energy may reduce our calculated event rate by a factor of 0.3. At higher energies it is even less. We have multipled our calculated event rates by this factor to obtain a conservative estimate.  As shown in Fig.\ref{optd} the mean free path of pair production in radio background is of the order of Mpc, hence it is not important for these gamma rays.  
 The observed diffuse UHECR spectrum has a spectral index $3$ between 1 PeV and 3 EeV \cite{apel2}. Assuming Fe nuclei and protons in 1:1 ratio we have calculated the expected event rates for this spectrum 
 in  $Km^2-yr$, shown by dotted line in Fig.4. We expect a significant number of gamma ray events at PeV energy.    
The number of electron and muon neutrinos, antineutrinos produced in hadronic interactions is comparable to the number of gamma rays produced at the same energy. It is much lower than the limit set by ANTARES \cite{antares} experiment at $7.5\times 10^{-8}(E_{\nu}/GeV)^{-2} GeV^{-1} cm^{-2} sec^{-1} sr^{-1}$ which corresponds to $3\times10^5$ events in $Km^2-yr$.  
 If there are point sources of very high energy gamma rays near the GC, PeV gamma rays may be detectable from them above the background calculated in this paper. A galactocentric very high energy gamma ray flux ($>$PeV) would be signature of EeV Galactic UHECRs. Thus the detection of PeV gamma rays would be useful to know about the origin of the UHECRs.  
\section{Conclusions}
While propagating in the Galaxy the diffuse UHECRs interact with ambient matter and radiations and produce secondary particles. Some of the UHECRs may be trapped in magnetic field in the GC region and produce galactocentric secondary gamma rays. If the UHECRs are coming from Galactic point sources, located near the GC, then we would expect a galactocentric flux of gamma rays from these point sources above the background calculated in this work. Thus PeV gamma ray astronomy would be very useful to know the highest energy of the Galactic cosmic rays. 
\section{Acknowledgment}
The author would like to thank R. Subrahmanyan for helpful discussions on the radio background and the referee for useful comments.

\end{document}